\documentclass[pra,aps,twocolumn,showpacs,floatfix]{revtex4}
\usepackage{tabularx}
\usepackage{amsmath}
\usepackage{amssymb}
\usepackage{graphicx}
\usepackage{dcolumn}
\usepackage{ulem}
\usepackage{bm}

\def\be{\begin{equation}}
\def\ee{\end{equation}}
\def\bea{\begin{eqnarray}}
\def\eea{\end{eqnarray}}
\def\bse{\begin{subequations}}
\def\ese{\end{subequations}}

\def\be{\begin{eqnarray}}
\def\ee{\end{eqnarray}}

\begin{document}

\title{Many-body Landau-Zener Transition in Cold Atom Double Well Optical
Lattices}
\author{Yinyin Qian}
\affiliation{Department of Physics, The University of Texas at Dallas, Richardson, Texas
75080 USA}
\affiliation{Department of Physics and Astronomy, Washington State University, Pullman,
Washington 99164 USA}
\author{Ming Gong}
\affiliation{Department of Physics, The University of Texas at Dallas, Richardson, Texas
75080 USA}
\affiliation{Department of Physics and Astronomy, Washington State University, Pullman,
Washington 99164 USA}
\author{Chuanwei Zhang}
\thanks{chuanwei.zhang@utdallas.edu}
\affiliation{Department of Physics, The University of Texas at Dallas, Richardson, Texas
75080 USA}
\affiliation{Department of Physics and Astronomy, Washington State University, Pullman,
Washington 99164 USA}

\begin{abstract}
Ultra-cold atoms in optical lattices provide an ideal platform for exploring
many-body physics of a large system arising from the coupling among a series
of small identical systems whose few-body dynamics is exactly solvable.
Using Landau-Zener (LZ) transition of bosonic atoms in double well optical
lattices as an experimentally realizable model, we investigate such few to
many body route by exploring the relation and difference between the small
few-body (in one double well) and the large many-body (in double well
lattice) non-equilibrium dynamics of cold atoms in optical lattices. We find
the many-body coupling between double wells greatly enhances the LZ
transition probability. The many-body dynamics in the double well lattice
shares both similarity and difference from the few-body dynamics in one and
two double wells. The sign of the on-site interaction plays a significant
role on the many-body LZ transition. Various experimental signatures of the
many-body LZ transition, including atom density, momentum distribution, and
density-density correlation, are obtained.
\end{abstract}

\maketitle

\section{Introduction}

\begin{figure}[t]
\centering
\includegraphics[width=3.3in]{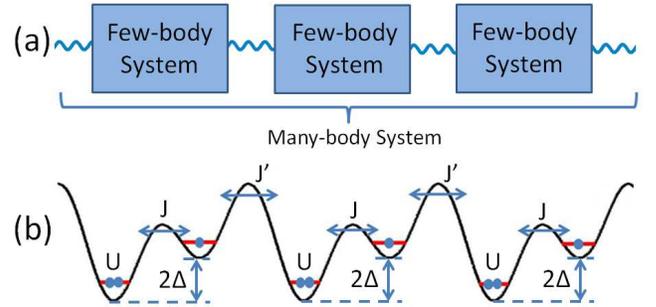}
\caption{(Color online). (a) Illustration of the route from the few-body
physics in identical small systems to the many-body physics in a large
system through coupling. (b) An example for (a): the LZ transition of cold
atoms in 1D double well optical lattices. The coupling between different
double wells (the small systems) is provided by the inter-well coupling $%
J^{\prime}$.}
\label{fig-lattice}
\end{figure}

Understanding many-body physics in strongly-correlated lattice models is
essential for the explanation of many important condensed matter phenomena,
such as the high temperature cuprate superconductivity \cite{Wen}. In this
context, ultra-cold atoms in optical lattices provide an ideal platform for
emulating numerous phenomena in solids because of their ability of
accurately implementing various lattice models without impurities, lattice
phonons, and other complications \cite%
{Bloch2005,Porto,Greiner02a,Jaksch98,Duan}. In solids, a large many-body
system may be composed of a series of small identical few-body systems,
therefore it would be natural and interesting to investigate how many-body
properties (\textit{e.g.} correlations) of the large system emerge or differ
from the few-body properties of the small systems (see Fig. \ref{fig-lattice}%
(a) for an illustration) \cite{Anderson}. Although such few to many body
route may provide a unique angle for understanding the underlying many-body
physics, its experimental realization is very challenging in solids. In
contrast, such route may be easily explored using the recent experimentally
realized cold atom double well optical lattices \cite{Strabley06,
Strabley07, Trotzky}, where the few-body dynamics in each double well can be
solved exactly, while the many-body physics emerges from the inter-well
coupling.

The double well lattices not only allow studying interesting many-body
ground states \cite{Ritt,Salger,ground1,ground2,ground3,ground4}, but also
the equilibrium and non-equilibrium dynamics of cold atoms after an
adiabatic or sudden change of the atom or lattice parameters \cite%
{Qian,Cramer}. Recently, the non-equilibrium dynamics in optical lattices
after a sudden quench has been investigated intensively \cite%
{Cramer,Rigol,Kollath,Manmana,Moeckel,Bernier11,Polkovnikov}. While the
dynamics for the adiabatic process is expected to follow the change of the
system Hamiltonian, the goal for studying the quench dynamics is to
understand the non-equilibrium physics and the relaxation to the equilibrium
steady states in the presence of many-body interactions. Although the two
limiting cases (adiabatic or sudden) have been widely studied, the
intermediate region, that is, the parameter variation with a finite rate,
has been largely unexplored \cite{Poletti,Bernier12,Natu,Clark}.

In this paper, we integrate these two important aspects (\textit{i.e.},
route from few to many body physics and non-equilibrium dynamics) for cold
atom optical lattices into one simple, but experimentally feasible model:
the many-body Landau-Zener (LZ) transition in a one-dimensional (1D) double
well optical lattice. In the LZ transition, the parameters of the
Hamiltonian vary with a finite rate (neither adiabatic nor sudden), and the
dynamics is naturally non-equilibrium. Furthermore, the dynamics of a single
atom in an isolated double well, a classical example of the LZ transition
\cite{Landau32}, is exactly solvable. Recently, the LZ transition has been
generalized (both theoretically and experimentally) to a BEC in a double
well (with $\sim $100 atoms per double well), where the dynamics is governed
by the mean-field nonlinear interaction \cite{Wu1,Wu2, Yuao,Kasztelan} and
the research focuses on the emergence of the loop structure in the energy
spectrum and the invalidity of the adiabaticity.\qquad\

In this paper we investigate the many-body LZ transition in a double well
optical lattice to explore the route from few-body to many-body
non-equilibrium dynamics. We mainly focus on the following issues: (i) the
LZ transition for a few interacting atoms (in contrast to hundreds of atoms
in the mean-field region \cite{Wu1,Wu2,Yuao}) in an isolated double well (%
\textit{i.e.}, without inter-well coupling); (ii) the collective LZ
transition of many atoms in the double well lattice with the inter-well
coupling, which has not been explored previously in the literature. Our goal
is to explore the difference as well as the relation between the few-body
physics in (i) and the many-body physics in (ii). We find that the onsite
interaction $U$ (repulsive or attractive) between atoms can strongly modify
the LZ transition in (i). For the repulsive interaction, there is an
oscillation of the LZ probability with respect to $U$. In the large $U$
limit, we derive an analytical expression for the LZ transition probability
using the independent crossing approximation. The inter-well tunneling that
couples different double wells can significantly increase the\ LZ transition
probability. While certain feature of the LZ transition process in a single
double well is still kept in the double well optical lattice, the coherent
oscillation of the transition probability in the positive $U$ region is
destroyed by the many-body tunneling between double wells. We show the
signature of the many-body LZ transition in various experimentally
measurable quantities such as the atom density and momentum distributions,
and the density-density correlation.

The rest of this paper is organized as follows: in section \ref{sec-ham} we
discuss the theoretical model of the experimentally realized double well
optical lattice. Then we study the few-body dynamics of the LZ transition in
a single double well in section \ref{sec-few}. We extend the dynamics of the
LZ transition to a coupled double well lattice in section \ref{sec-many} by
switching on the inter-well coupling. In section \ref{sec-ob} we discuss
possible experimental signatures of many-body LZ transitions. Section \ref%
{sec-con} is a summary.

\section{The Hamiltonian}

\label{sec-ham}

The 1D double well lattice, schematically shown in Fig. \ref{fig-lattice}%
(b), has been realized in many experiments by superimposing two optical
lattices with different wavelengths \cite{Strabley06, Strabley07, Trotzky,
Ritt, Salger}. The dynamics along the other two dimensions is frozen to the
ground states using optical lattices with high lattice potential depths.
Within the tight-binding approximation, the dynamics of atoms in the 1D
double well lattice is described by the Bose-Hubbard model \cite{Jaksch98},
\begin{eqnarray}
H &=&-\sum\nolimits_{i}\left( Ja_{2i-1}^{\dag }a_{2i}+J^{\prime
}a_{2i}^{\dag }a_{2i+1}+c.c.\right)  \notag \\
&&+\sum\nolimits_{i}\left( -1\right) ^{i+1}\Delta n_{i}+Un_{i}\left(
n_{i}-1\right) ,  \label{Ham1}
\end{eqnarray}%
where $a_{i}$ ($a_{i}^{\dagger }$) are the annihilation (creation) operators
for bosons at the lattice site $i$, $n_{i}=a_{i}^{\dagger }a_{i}$, $J$ ($%
J^{\prime }$) represents the intra-well (inter-well) tunneling, $2\Delta
=2\gamma t$ is the chemical potential difference between two neighboring
lattice sites in a single double well, which varies with the time $t$ with
the sweeping rate $\gamma $. $U$ is the on-site interaction strength between
atoms. In experiments \cite{Strabley06, Strabley07, Trotzky}, the inter- and
intra-well tunnelings $J$ and $J^{\prime }$ can be adjusted independently by
careful control of the intensities of the two optical lattice laser beams,
the chemical potential $\Delta $ can be tuned by shifting one laser beam
with respect to the other, and the on-site interaction $U$ between atoms can
be changed using the Feshbach resonance. Generally the inter-well barrier,
see Fig. \ref{fig-lattice}(a) is larger than the intra-well barrier, i.e., $%
J>J^{\prime }$.

The time-dependent dynamics of atoms governed by the Hamiltonian (\ref{Ham1}%
) is solved numerically. Because the dimensionality of the Hamiltonian
increases exponentially with respect to the lattice size, we simulate the
dynamics of this system using the recently developed \textit{time-evolving
block decimation} (TEBD) algorithm \cite{Vidal03,Vidal04,Vidal07,TEBD} with
an open boundary condition, which is a powerful tool for studying lattice
dynamics in one dimension with only nearest neighbor tunneling/interaction
where the entanglement of the system is small. In our TEBD simulation, we
choose the Schmidt number $\xi =20$ with at most 4 atoms at each lattice
site. We have confirmed the convergence of our numerical program for the
parameters used in the paper by comparing the results with that using a
larger Schmidt number. For a small lattice size, we also study the dynamics
using the Runge-Kutta method by including all possible configurations and
find excellent agreement with the TEBD method. Henceforth, the energy unit
is chosen as the intra-well tunneling $J$, and the corresponding time unit
is $\hbar J^{-1}$.

\begin{figure}[t]
\centering
\includegraphics[width=3.3in]{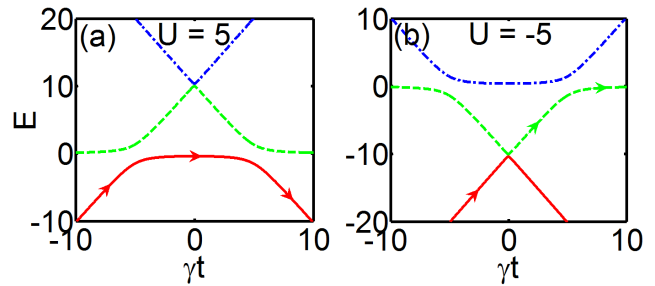}
\caption{(Color online). Plot of the instantaneous energy levels in a single
double well with two atoms. The level crossing depends strongly on the sign
of $U$. The arrow indicates the time evolution of the quantum state for a
relatively small sweeping rate $\protect\gamma$.}
\label{fig-energy}
\end{figure}

\section{Few-body dynamics in one double well}

\label{sec-few}

For cold atoms in isolated double wells without inter-well tunneling (i.e., $%
J^{\prime }=0$), the probability for the atoms remaining on the same state
after they pass through the LZ transition regime is known to be $\exp \left(
-\pi J^{2}/\gamma \right) $ \cite{Landau32} for $U=0$. In the presence of
the on-site interaction $U$ for several atoms in one double well, the LZ
transition can be dramatically different because the interaction shifts the
energy levels for atoms. For simplicity, we illustrate the essential physics
using two atoms ($N=2$) in one double well. Under the number state basis $%
\left\{ \left\vert 2_{L}0_{R}\right\rangle ,\left\vert
1_{L}1_{R}\right\rangle ,\left\vert 0_{L}2_{R}\right\rangle \right\} $, the
Hamiltonian can be written as
\begin{equation}
H_{2}=%
\begin{pmatrix}
2\gamma t+2U & -\sqrt{2}J & 0 \\
-\sqrt{2}J & 0 & -\sqrt{2}J \\
0 & -\sqrt{2}J & -2\gamma t+2U%
\end{pmatrix}%
,  \label{eq-h2}
\end{equation}%
where $L$ ($R$) represents the left (right) well. The instantaneous
eigenenergy levels at the time $t$ are plotted in Fig. \ref{fig-energy}.
With the large $\left\vert U\right\vert \gg J$, the two large anticrossings
at $t\neq 0$ with the gap $\sim J$ yield direct LZ transitions between two
quantum states that differ by exactly one atom, while the tiny anticrossing
at $t=0$ with the gap $\sim J^{2}/U$ corresponds to the indirect second
order transition process, which has a negligible contribution to the total
LZ transition probability when the sweeping rate is faster than the time
scale determined by this gap. Assuming initially the atoms are prepared on
the ground state $\left\vert 2_{L}0_{R}\right\rangle $ for $t\rightarrow
-\infty $, we see for $U>0$, all atoms can be transferred from the left to
the right well at $t\rightarrow +\infty $ when $\gamma $ is small (the arrow
in Fig. \ref{fig-energy}(a)). While for $U\ll -J$ at most one atom can be
transferred for a relatively slow sweeping rate $\gamma $ (but still fast
enough such that the tiny gap at $t=0$ can be neglected). For a very small
sweeping rate $\gamma $ (thus the gap at $t=0$ cannot be neglected), the
dynamics is still adiabatic and all atoms can be transferred to the right
well, as expected. For the atom number larger than two ($N>2$), the physical
picture shown in Fig. \ref{fig-energy} is still similar. Generally, for $%
U\gg J$ we observe $N$ different LZ transitions, while for $U\ll -J$ only
one direct LZ transition can be found (hence at most one particle can be
transferred from left to right if the sweeping rate is faster than the
indirect transition gap). Finally, assuming the $i$-th LZ transition occurs
at the time $t=t_{i}$ where the anticrossing between $|i,N-i\rangle $ and $%
|i+1,N-i-1\rangle $ takes place. At time $t_{i}$, the two states have the
same energy, that is, $Ui(i-1)+U(N-i)(N-i-1)+\gamma
t_{i}(i-(N-i))=Ui((i+1)-1)+U(N-(i+1))(N-(i+1)-1)+\gamma
t_{i}((i+1)-(N-(i+1)))$, yielding $t_{i}=((N-1)U-2iU)/\gamma $. Therefore
the time interval between two adjacent direct LZ transitions is exactly the
same $\delta t=t_{i}-t_{i+1}=2U/\gamma $.

The exact time-dependent dynamics of the Hamiltonian (\ref{eq-h2}) is very
complex and cannot be expressed using simple analytical equations. However,
simple analytical expressions for the remaining number of atoms in the left
well can be derived in certain limit using the $S$-matrix method developed
by Brundobler and Elser \cite{Brundobler93}, which has been successfully
applied to other similar models \cite%
{Sinitsyn02,Dobrescu06,Volkov04,Sinitsyn04,Volkov05,Demkov68,Demkov01,Pokrovsky02, Kayanuma84,Pokrovsky04, Sinitsyn03,Pokrovsky07}%
. With the $S$-matrix method, we find
\begin{equation}
N_L = \left\{
\begin{array}{ll}
1 + e^{-2\pi J^2/\gamma} & \quad \text{if $U \gg J$} \\
e^{-2\pi J^2/\gamma}(3 - e^{-2\pi J^2/\gamma}) & \quad \text{if $U \ll -J$}
\\
2 e^{-2\pi J^2/\gamma} & \quad $U = 0$ \\
&
\end{array}
\right.
\end{equation}
using the independent crossing approximation (ICA) \cite{Brundobler93} that
is valid in these limits, as shown in Fig. \ref{fig-energy}. We have
verified that the above analytical results agree well with our numerical
results. For $U=0$, the result reduces to the standard LZ formula \cite%
{Landau32}, multiplied by the total number of atoms.

\section{Many-body dynamics in double well lattice}

\label{sec-many}

\begin{figure}[t]
\centering
\includegraphics[width=2.4in]{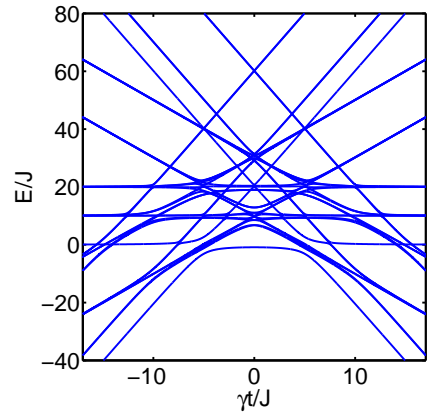}
\caption{(Color online). Plot of the instantaneous energy levels in two
coupled double wells with two atoms in each double well. $J^{\prime }= 0.5J$%
, $U = 5.0 J$, $M = 2$. The strong crossings and anticrossings between
different quantum states make the independent crossing approximation invalid
in the multiple coupled double wells.}
\label{fig-energy2wells}
\end{figure}

Now we switch on the coupling $J^{\prime }$ between double wells to
investigate how $J^{\prime }$ and $U$ influence the LZ transitions. To
illustrate the essential physics in the large lattice, we first consider an
isolated system with only two coupled double wells, each of which contains $%
N=2$ atoms. Similar as the Hamiltonian (\ref{eq-h2}) for a single double
well, we rewrite the Hamiltonian (\ref{Ham1}) to a $35\times 35$ matrix form
under the number state basis $\left\{ \left\vert
i_{1}i_{2}i_{3}i_{4}\right\rangle \right\} $, where $i_{l}=\left\{
0,1,2,3,4\right\} $ represents the number of atoms at the site $l$ with the
constraint $i_{1}+i_{2}+i_{3}+i_{4}=4$. To gain insight into this problem,
we plot the instantaneous energy levels for $U>0$ in Fig. \ref%
{fig-energy2wells}. We see the ICA \cite{Brundobler93} is invalid even for a
very large $U$, therefore the analytical formula for the number of atoms in
the left well cannot be derived anymore. The same physical picture holds
when multiple double wells in the lattice are coupled by switching on the
inter-well coupling $J^{\prime }$.

\begin{figure}[b]
\centering
\includegraphics[width=2.7in]{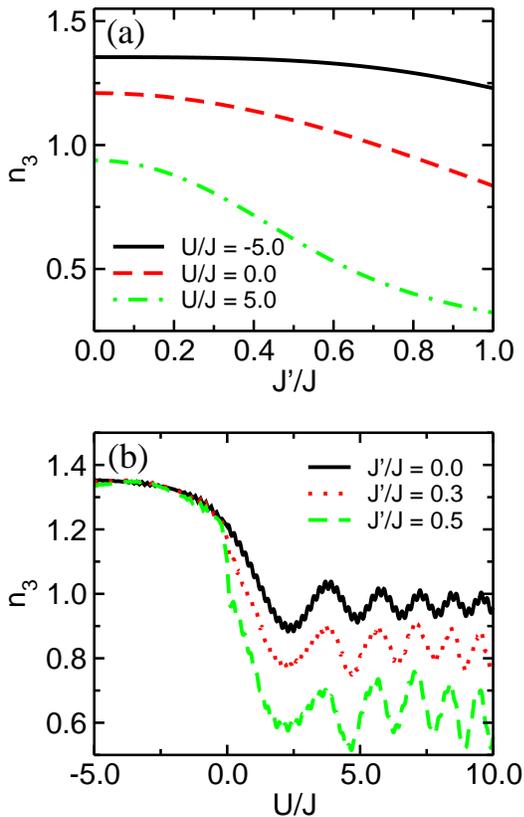}
\caption{(Color online). Plot of the atom number $n_{3}$ in the third site
versus (a) $J^{\prime }$ and (b) $U$ for two coupled double wells from the
Runge-Kutta simulation. $\protect\gamma =2\protect\pi $.}
\label{fig-2wells}
\end{figure}

We numerically solve the dynamics of atoms in two double wells using the
fourth-order Runge-Kutta method to obtain the exact dynamics using an open
boundary condition. The quantum state of the system is initially prepared at
$|2_{L}0_{R}\rangle \otimes |2_{L}0_{R}\rangle $ (the ground state at $%
t\rightarrow -\infty $), and the number of atoms in the third site $n_{3}$
for a large positive $t$ (effectively $t\rightarrow +\infty $), i.e., the
number of atoms remaining in the left well, is calculated. Note that a small
$n_{3}$ means a large LZ transition probability. In Fig. \ref{fig-2wells}%
(a), we plot $n_{3}$ as a function of $J^{\prime }$ with the finite $U$. $%
n_{3}$ decreases monotonically with the increasing $J^{\prime }$ for all
interaction regions, which indicates the LZ transition probability is
enhanced by $J^{\prime }$. This can be intuitively understood from the fact
that when the inter-well tunneling is switched on, the atoms in the third
site can tunnel to not only the fourth site, but also the second site.

In Fig. \ref{fig-2wells}(b), we plot $n_{3}$ with respect to $U$ for finite $%
J^{\prime }$. We see a strong dependence of $n_{3}$ on the sign of $U$. For
a large attractive $U$, $n_{3}$ is large and the LZ transition probability
is small because atoms are bound together for tunneling by the large
attractive interaction, as shown in Fig. \ref{fig-energy}(b). The inter-well
tunneling $J^{\prime }$ barely modifies the transition probability. While in
the repulsive $U$ region, $n_{3}$ decreases quickly with the increasing $U$,
but saturates at the large $U$ limit. In the large $U$ region, $n_{3}$
oscillates periodically with $U$ even for $J^{\prime }=0$. Note that there
are two direct LZ transitions for two atoms in a double well. After passing
the minimum gap point of the first LZ transition, $n_{3}$ still oscillates
periodically with time. Since the time interval between two adjacent direct
LZ transitions is $\delta t=2U/\gamma $, it is expected that $n_{3}$ before
the second LZ transition may oscillate with $U$, leading to the final
oscillation dependence of $n_{3}$ on $U$ at $t\rightarrow +\infty $. The
nonzero $J^{\prime }$ reduces $n_{3}$ significantly by enhancing the LZ
transition probability. However, the positions of the peaks and valleys of
the oscillation are the same for different $J^{\prime }$. Note that general
analytical perturbation methods (with $J^{\prime }$ as a small parameter) do
not work for the coupled LZ transition because of the non-adiabaticity and
the time-dependence of the system.

\begin{figure}[tbp]
\centering
\includegraphics[width=2.7in]{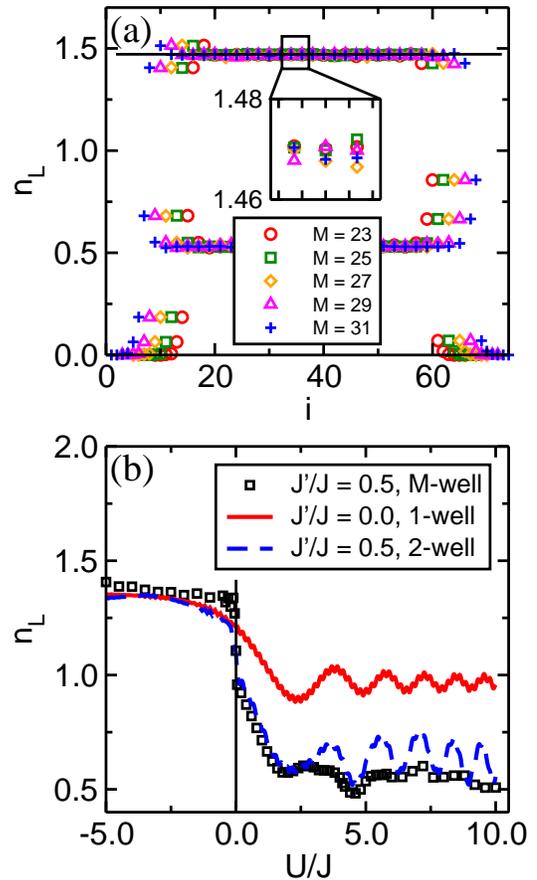}
\caption{(Color online). (a) Plot of the atom density distribution $n_{i}$
for different lattice sizes from the TEBD simulation. $M$ is the number of
double wells. $J^{\prime }=0.5J$, $U=5.0J$, $\protect\gamma =2\protect\pi $.
An enlarged plot at the center of the lattice is shown in the inset. (b)
Plot of the number of atoms at the left site $n_{L}$ of the central double
well of the lattice with respect to $U$. $M=25$, and $\protect\gamma =2%
\protect\pi $.}
\label{fig-L}
\end{figure}

With the knowledge of the LZ transition in single and coupled double wells,
we now study the many-body LZ transitions in a lattice with many ($\sim 30$)
double wells using the TEBD algorithm. Note that because of the
exponentially large dimensionality of the lattice system, the fourth-order
Runge-Kutta method (as well as any exact diagonalization method) is not
practical. The initial wavefunction is assumed to be $|0_{L}0_{R}\rangle
^{\otimes N_{L}}\otimes |2_{L}0_{R}\rangle ^{\otimes M}\otimes
|0_{L}0_{R}\rangle ^{\otimes N_{R}}$, where $M$ is the number of occupied
double wells. $N_{L}$ ($N_{R}$) is the number of unoccupied double wells at
the left (right) edge of the lattice to eliminate the effects of the open
boundary condition. With the finite $J^{\prime }$, the atoms may diffuse to
the left and right edges in the time scale $\sim 1/J^{\prime }$. For a
relatively large $\gamma $, such diffusion does not affect the LZ transition
at the center of the lattice, as we show below. In the numerical simulation,
we choose $N_{L}=N_{R}=3$, which are sufficient for a wide range of $\gamma $%
.

To check the effects of the atom diffusion on the LZ transition, we plot the
number of atoms (at $t\rightarrow +\infty $) in each lattice site in Fig. %
\ref{fig-L}(a) for $\gamma =2\pi $ and different $M$. Although the atom
diffusion at the edge of the lattice is significant, the atom number
fluctuation in the central part of the lattice is generally very small. For
instance, the difference between the final atom number at the center of the
lattice is smaller than $0.002$ for two different lattice sizes $M=23$ and $%
31$. Our numerical results show that when $M>20$ the LZ transition in the
central part of the lattice can be a good approximation for that in an
infinite large lattice $M\rightarrow \infty $.

We calculate the number of atoms $n_{L}$ (at $t\rightarrow +\infty $) in the
left site of the double well at the center of the lattice with $M=25$, and
plot it as a function of $U$ in Fig. \ref{fig-L}(b). To compare the results
with the few-body physics in the small system, we also present the results
for single and two coupled double wells in the same figure. When $U<0$, $%
n_{L}$ is generally larger than one because of the reduced LZ transition
probability by the attractive interaction. $n_{L}$ is almost the same as
that in coupled double wells, which indicates $J^{\prime }$ does not modify
the LZ transition significantly for the attractive interaction. For $U>0$, $%
n_{L}$ is generally smaller than 1. Although in a large quantum system the
instantaneous eigenenergy levels become very complex, we still observe the
oscillation of $n_{L}$ with respect to $U$. However, the perfect oscillation
in the coupled double wells is smeared by the interference effects in
lattices, leaving only one big dip at $U\simeq 4.5J$. This is very different
from the coherent oscillation in one and two coupled double wells.

\begin{figure}[t]
\centering
\includegraphics[width=0.95\linewidth]{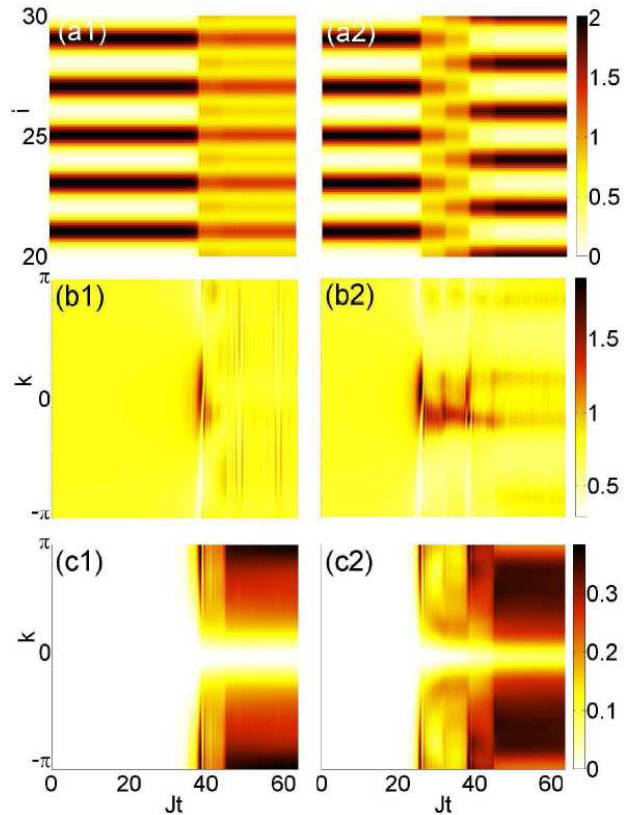}
\caption{(Color online). Plot of the atom density $n_{i}(t)$ (a1, a2),
momentum distribution $n(k,t)$ (b1, b2), and density-density correlation $%
s(k,t)$ (c1, c2) from TEBD numerics. $J^{\prime }=0.5J$. The left column: $%
U=-5J$; the right column: $U=5J$. $\protect\gamma =2\protect\pi $. In all
panels, the dark contrast lines correspond to places for the LZ transitions
(see main text for more details).}
\label{fig-nk}
\end{figure}

\section{Experimental observation}

\label{sec-ob}

The many-body LZ transition can be observed using several different methods.

(i) The number of atoms in the left or right wells of the lattice (e.g., $%
n_{L}$ in Fig. \ref{fig-L}) can be measured using the band mapping method
\cite{Trotzky}. In Figs. \ref{fig-nk}(a), we plot the atom density
distribution during the LZ transition in the central part of the lattice. We
see there is only one LZ transition for $U\ll -J$, while several LZ
transitions for $U\gg J$.

(ii) When one atom is transferred from one site to the neighboring site,
there should be a change of the atom momentum distribution
\begin{equation}
n(k,t)=\sum_{i,j}e^{ik\left( i-j\right) }\langle \psi |a_{i}^{\dagger
}a_{j}|\psi \rangle ,  \label{eq-md}
\end{equation}%
which is plotted in Figs. \ref{fig-nk}(b) and can be measured using the
time-of-flight image. Here $\psi \left( t\right) $ is the wavefunction
obtained from the TEBD numerical simulation. Note that the actual momentum
distribution may oscillate rapidly after the LZ transition, which may not be
observable in experiments. Here, we take the average momentum distribution
in a small time interval\ 0.9$\hbar J^{-1}$ in Figs. \ref{fig-nk}(b). We see
the atom momentum distributions $n(k,t)$ have peaks around $k=0$ during the
LZ transitions. There is a one-to-one correspondence between the density and
momentum changes in the LZ transition, therefore the momentum peaks around $%
k=0$ (the dark lines in Figs. \ref{fig-nk}(b)) appear once for $U<0$ (one LZ
transition) and several times for $U>0$.

(iii) The density-density correlation
\begin{equation}
s(k,t)=\sum_{i,j}e^{ik\left( i-j\right) }\left[ \langle \psi
|n_{i}n_{j}|\psi \rangle -\langle \psi |n_{i}|\psi \rangle \langle \psi
|n_{j}|\psi \rangle \right] ,  \label{eq-dc}
\end{equation}%
which can be measured using the Bragg scattering \cite{Brag}, shows similar
features as $n_{L}$ because $n_{i}n_{j}$ should also have a sudden change in
the LZ transition process. Similarily, there are different number of dark
lines in Figs. \ref{fig-nk}(c) for $U<0$ and $U>0$. Note that $n(k,t)$ shows
a strong asymmetry about $k=0$ because the momentum is transferred along a
specific direction during the LZ transition. In contrast, the
density-density correlation, due to its definition, is exactly symmetric
about $k=0$. In experiments, the number of atoms in the left well $n_{L}$
and the densiy-density correlation $s(k,t)$ can also be measued using the
single site detection \cite{Greiner,Bloch3,Bloch4}.

Experimentally, the 1D double well lattice has been realized for $^{87}$Rb
atoms by superimposing two optical lattices with the wavelength $\lambda
_{1}=\lambda _{2}/2$ and the corresponding lattice potential depths $V_{1}$
and $V_{2}$ \cite{Trotzky}. The motion of atoms along the other two
dimensions are frozen by two additional regular lattices with the potential
depth $V_{\bot }=30E_{R}$, where $E_{R}=h^{2}/2m\lambda ^{2}=2\pi \times 3.7$
KHz is the atom recoil energy. By tuning $V_{1}$ and $V_{2}$, we can set the
intra-well tunneling energy $J\sim \hbar \times 2\pi \times 200$ Hz. The
typical onsite interaction strength in the optical lattice is $U\sim $ $%
\hbar \times 2\pi \times 1$ KHz, but can be tuned using Feshbach resonance
\cite{Chin10}. The chemical potential $\Delta $ can be tuned by shifting one
laser beam with respect to the other \cite{Trotzky}. The typical LZ
transition time is $\sim $50$\hbar /J\sim 40$ ms, which can be easily
achieved in experiments. Taking into account of the symmetry of the energy
levels for the repulsive and attractive interactions (Figs. \ref{fig-energy}%
(a) and \ref{fig-energy}(b)), we can use the repulsive interaction ($U>0$)
to emulate the LZ transition with the attractive interaction ($U<0$) by
initially preparing the atoms in the excited state (the upper branch in Fig. %
\ref{fig-energy}(a)), similar as the method used in a recent experiment \cite%
{Trotzky}.

\section{Conclusion}

\label{sec-con}

In summary, we study the many-body effects in the LZ transition using
ultra-cold bosonic atoms in 1D double well optical lattices. The dynamics of
such non-adiabatic systems is obtained through large scale numerical
simulations using the TEBD algorithm. Our results are important for
understanding not only the route from few to many body physics when
individual small systems are coupled to form a large system through the
many-body coupling, but also the non-equilibrium dynamics in optical
lattices when the system parameters vary with a finite rate (neither
adiabatic nor sudden).

\emph{Acknowledgement:} We thank Yu-Ao Chen for helpful discussion. This
work is supported by ARO (W911NF-09-1-0248), DARPA-YFA (N66001-10-1-4025),
and NSF (PHY-1104546).

\end{document}